\documentclass[aps,preprint,tightenlines]{revtex4}%
\usepackage{amsmath}
\usepackage{graphicx}
\usepackage{amsfonts}
\usepackage{amssymb}%
\begin{document}
\preprint{ }
\title{Variational path-integral treatment of a translation invariant many-polaron system}
\author{F. Brosens}
\author{S. N. Klimin}
\author{J. T. Devreese}
\altaffiliation{Permanent address: Department of Theoretical Physics, State University of
Moldova, str. A. Mateevici 60, MD-2009 Kishinev, Republic of Moldova.}

\altaffiliation{Also at Technische Universiteit Eindhoven, P. B. 513, 5600 MB Eindhoven, The Netherlands.}

\affiliation{Theoretische Fysica van de Vaste Stoffen (TFVS), Universiteit Antwerpen,
B-2610 Antwerpen, Belgium}

\begin{abstract}
A translation invariant $N$-polaron system is investigated at arbitrary
electron-phonon coupling strength, using a variational principle for path
integrals for identical particles. An upper bound for the ground state energy
is found as a function of the number of spin up and spin down polarons, taking
the electron-electron interaction and the Fermi statistics into account. The
resulting addition energies and the criteria for multipolaron formation are discussed.

\end{abstract}
\volumeyear{year}
\volumenumber{number}
\issuenumber{number}
\eid{identifier}
\date{January 24, 2006}
\startpage{1}
\endpage{100}
\maketitle

\section{Introduction}

Thermodynamic and optical properties of interacting many-polaron systems are
intensely investigated, because they might play an important role in physical
phenomena in high-$T_{c}$ superconductors (see, e.g.,
Refs.~\cite{alex,Polarons} and references therein). In particular, numerous
experiments on the infrared optical absorption of high-$T_{c}$ materials (see,
e.g., Refs. \cite{Genzel89,falck1,calva1b,QQ4,Lupi1999,Hartinger2004}) reveal
features which are associated with large polarons
\cite{Lupi1999,Hartinger2004,jt1}.

For the case of weak electron-phonon coupling strength, a suitable variational
approximation to the ground state energy of an interacting many-polaron gas
was already developed in \cite{LDB77}, using a many-body canonical
transformation for fermions in interaction with a phonon field. The static
structure factor of the electron gas is the key ingredient of this theory.
Based on the approach of Ref. \cite{LDB77}, a many-body theory for the optical
absorption at a gas of interacting polarons was developed \cite{TD2001}. The
resulting optical conductivity turns out to be in fair agreement with the
experimental \textquotedblleft\emph{d} band\textquotedblright\ by Lupi
\emph{et al.} \cite{Lupi1999} in the optical-absorption spectra of cuprates.

At arbitrary electron-phonon coupling strength, the many-body problem
(including electron-electron interaction and Fermi statistics) in the
$N$-polaron theory is not well developed. Within the random-phase
approximation, the optical absorption of an interacting polaron gas was
studied in Ref. \cite{Iadonisi}, taking over the variational parameters of
Feynman's polaron model \cite{Feynman55}, which however are derived for a
single polaron without many-body effects. For a dilute arbitrary-coupling
polaron gas, the equilibrium properties \cite{Q1,Q2} and the optical response
\cite{Fr} have been investigated using the path-integral approach taking into
account the electron-electron interaction but neglecting the Fermi statistics.
Recently, the formation of many-polaron clusters was investigated in Ref.
\cite{EPJ-B41-163} using the Vlasov kinetic equations \cite{Vlasov}. However,
also this approach does not take into account the Fermi statistics of
electrons, and therefore it is only valid for sufficiently high temperatures.

The path integral treatment \cite{PRE96,PRE97,SSC99} of the quantum statistics
of indistinguishable particles (bosons or fermions) provides a sound basis for
including the many-body effects in a system of interacting polarons
\cite{NoteKleinert}. This approach was used \cite{MPSSC,MPPhysE,MPQD2004} for
calculating the ground state energy and the optical conductivity spectra at
arbitrary electron-phonon coupling strength for a finite number of interacting
polarons in a parabolic confinement potential. However, the translation
invariant polaron gas was not yet investigated within this approach.

In the present work, the ground-state properties of a translation invariant
$N$-polaron system are theoretically studied in the framework of the
variational path-integral method for identical particles, using a further
development of the model introduced in Refs. \cite{MPSSC,MPPhysE,MPQD2004}. In
Sec. II, the variational path-integral method and the chosen model system are
described. In Sec. III, we discuss the numerical results for the ground-state
energy of a translation invariant $N$-polaron system. Sec. IV is a summary of
the obtained results with conclusions.

\section{Variational path-integral method for a $N$-polaron system}

\subsection{The many-polaron system}

In order to describe a many-polaron system, we start from the translation
invariant $N$-polaron Hamiltonian
\begin{equation}
H=\sum_{j=1}^{N}\frac{\mathbf{p}_{j}^{2}}{2m}+\frac{1}{2}\sum_{j=1}^{N}%
\sum_{l=1,\neq j}^{N}\frac{e^{2}}{\epsilon_{\infty}\left\vert \mathbf{r}%
_{j}\mathbf{-r}_{l}\right\vert } +\sum_{\mathbf{k}}\hbar\omega_{\mathrm{LO}%
}a_{\mathbf{k}}^{\dagger}a_{\mathbf{k}}+\left(  \sum_{j=1}^{N}\sum
_{\mathbf{k}}V_{k}a_{\mathbf{k}}e^{i\mathbf{k\cdot r}_{j}}+H.c.\right)  ,
\end{equation}
where $m$ is the band mass, $e$ is the electron charge, $\omega_{\mathrm{LO}}$
is the longitudinal optical (LO) phonon frequency, and $V_{k}$ are the
amplitudes of the Fr\"{o}hlich electron-LO-phonon interaction%
\begin{equation}
V_{k}=i\frac{\hbar\omega_{\mathrm{LO}}}{k}\left(  \frac{4\pi\alpha}{V}\right)
^{1/2}\left(  \frac{\hbar}{2m\omega_{\mathrm{LO}}}\right)  ^{1/4},\quad
\alpha=\frac{e^{2}}{2\hbar\omega_{\mathrm{LO}}}\left(  \frac{2m\omega
_{\mathrm{LO}}}{\hbar}\right)  ^{1/2}\left(  \frac{1}{\epsilon_{\infty}}%
-\frac{1}{\epsilon_{0}}\right)  ,
\end{equation}
of course with the electron-phonon coupling constant $\alpha>0,$ the
high-frequency dielectric constant $\epsilon_{\infty}>0$ and the static
dielectric constant $\epsilon_{0}>0,$ and consequently
\begin{equation}
\frac{e^{2}}{\epsilon_{\infty}}>\hbar\left(  \frac{2\hbar\omega_{\mathrm{LO}}%
}{m}\right)  ^{1/2}\alpha\Longleftrightarrow\alpha\sqrt{2}<\left(
\frac{H^{\ast}}{\hbar\omega_{\mathrm{LO}}}\right)  ^{1/2}\equiv U,
\label{units}%
\end{equation}
which is an important physical condition on the relative strength of the
Coulomb interaction as compared to the electron-phonon coupling, as stressed
in the earlier bipolaron work \cite{VPD91}. In the expression (\ref{units}),
$H^{\ast}$ is the effective Hartree
\begin{equation}
H^{\ast}=\frac{e^{2}}{\epsilon_{\infty}a_{B}^{\ast}},\quad a_{B}^{\ast}%
=\frac{\hbar^{2}}{me^{2}/\epsilon_{\infty}}%
\end{equation}
where $a_{B}^{\ast}$ is the effective Bohr radius. The partition function of
the system can be expressed as a path integral over all electron and phonon
coordinates. The path integral over the phonon variables can be calculated
analytically \cite{Feynman}. Feynman's phonon elimination technique for this
system is well known and leads to the partition function, which is a path
integral over the electron coordinates only:%
\begin{equation}
Z=\left(  \prod_{\mathbf{k}}\frac{e^{\frac{1}{2}\beta\hbar\omega_{\mathrm{LO}%
}}}{2\sinh\frac{1}{2}\beta\hbar\omega_{\mathrm{LO}}}\right)  \oint
e^{S}\mathcal{D}\mathbf{\bar{r}}%
\end{equation}
where $\mathbf{\bar{r}=}\left\{  \mathbf{r}_{1},\cdots,\mathbf{r}_{N}\right\}
$ denotes the set of electron coordinates, and $\oint\mathcal{D}%
\mathbf{\bar{r}}$ denotes the path integral over all the electron coordinates,
integrated over equal initial and final points, i.e.
\[
\oint e^{S}\mathcal{D}\mathbf{\bar{r}}\equiv\int d\mathbf{\bar{r}}%
\int_{\mathbf{\bar{r}}\left(  0\right)  =\mathbf{\bar{r}}}^{\mathbf{\bar{r}%
}\left(  \beta\right)  =\mathbf{\bar{r}}}e^{S}\mathcal{D}\mathbf{\bar{r}%
}\left(  \tau\right)  .
\]
Throughout this paper, imaginary time variables are used. The effective action
for the $N$-polaron system is retarded and given by%
\begin{align}
S  &  =-\int_{0}^{\beta}\left(  \frac{m}{2}\sum_{j=1}^{N}\left(
\frac{d\mathbf{r}_{j}\left(  \tau\right)  }{d\tau}\right)  ^{2}+\frac{1}%
{2}\sum_{j=1}^{N}\sum_{l=1,\neq j}^{N}\frac{e^{2}}{\epsilon_{\infty}\left\vert
\mathbf{r}_{j}\left(  \tau\right)  \mathbf{-r}_{l}\left(  \tau\right)
\right\vert }\right)  d\tau\nonumber\\
&  +\frac{1}{2}\int_{0}^{\beta}\int_{0}^{\beta}\sum_{j,l=1}^{N}\sum
_{\mathbf{k}}\left\vert V_{k}\right\vert ^{2}e^{i\mathbf{k\cdot}\left(
\mathbf{r}_{j}\left(  \tau\right)  -\mathbf{r}_{l}\left(  \sigma\right)
\right)  }\frac{\cosh\hbar\omega_{\mathrm{LO}}\left(  \frac{1}{2}%
\beta-\left\vert \tau-\sigma\right\vert \right)  }{\sinh\frac{1}{2}\beta
\hbar\omega_{\mathrm{LO}}}d\sigma d\tau. \label{eq:Spol}%
\end{align}
Note that the electrons are fermions. Therefore the path integral for the
electrons with parallel spin has to be interpreted as the required
antisymmetric projection of the propagators for distinguishable particles.

We below use units in which $\hbar=1$, $m=1$, and $\omega_{\mathrm{LO}}=1$.
The units of distance and energy are thus the effective polaron radius
$\left[  \hbar/\left(  m\omega_{\mathrm{LO}}\right)  \right]  ^{1/2}$ and the
LO-phonon energy $\hbar\omega_{\mathrm{LO}}$.

\subsection{Variational principle}

For distinguishable particles, it is well known that the Jensen-Feynman
inequality \cite{Feynman55,Feynman} provides a lower bound on the partition
function $Z$ (and consequently an upper bound on the free energy $F$)%
\begin{equation}
Z=\oint e^{S}\mathcal{D}\mathbf{\bar{r}}=\left(  \oint e^{S_{0}}%
\mathcal{D}\mathbf{\bar{r}}\right)  \left\langle e^{S-S_{0}}\right\rangle
_{0}\geq\left(  \oint e^{S_{0}}\mathcal{D}\mathbf{\bar{r}}\right)
e^{\left\langle S-S_{0}\right\rangle _{0}}\text{ with }\left\langle
A\right\rangle _{0}\equiv\frac{\oint A\left(  \mathbf{\bar{r}}\right)
e^{S_{0}}\mathcal{D}\mathbf{\bar{r}}}{\oint e^{S_{0}}\mathcal{D}%
\mathbf{\bar{r}}}, \label{eq:JensenFeynman}%
\end{equation}%
\begin{equation}
e^{-\beta F}\geq e^{-\beta F_{0}}e^{\left\langle S-S_{0}\right\rangle _{0}%
}\Longrightarrow F\leq F_{0}-\frac{\left\langle S-S_{0}\right\rangle _{0}%
}{\beta} \label{VarIneq}%
\end{equation}
for a system with real action $S$ and a real trial action $S_{0}.$The
many-body extension (Ref.\thinspace\cite{PRE96}, p. 4476) of the
Jensen-Feynman inequality, discussed in more detail in Ref.\thinspace
\cite{NoteKleinert}, requires (of course) that the potentials are symmetric
with respect to all particle permutations, and that the exact propagator as
well as the model propagator are defined on the same state space. This means
that both the exact and the model propagator are antisymmetric for fermions
(symmetric for bosons) at any time. The path integrals in
(\ref{eq:JensenFeynman}) thus have to be interpreted in terms of an
antisymmetric state space. Within this interpretation we consider the
following generalization of Feynman's trial action%
\begin{align}
S_{0}  &  =-\int_{0}^{\beta}\left(  \frac{1}{2}\sum_{j=1}^{N}\left(
\frac{d\mathbf{r}_{j}\left(  \tau\right)  }{d\tau}\right)  ^{2}+\frac
{\omega^{2}+w^{2}-v^{2}}{4N}\sum_{j,l=1}^{N}\left(  \mathbf{r}_{j}\left(
\tau\right)  \mathbf{-r}_{l}\left(  \tau\right)  \right)  ^{2}\right)
d\tau\nonumber\\
&  -\frac{w}{8}\frac{v^{2}-w^{2}}{N}\sum_{j,l=1}^{N}\int_{0}^{\beta}\int
_{0}^{\beta}\left(  \mathbf{r}_{j}\left(  \tau\right)  -\mathbf{r}_{l}\left(
\sigma\right)  \right)  ^{2}\frac{\cosh w\left(  \frac{1}{2}\beta-\left\vert
\tau-\sigma\right\vert \right)  }{\sinh\frac{1}{2}\beta w}d\sigma
d\tau\label{eq:S0}%
\end{align}
with the variational frequency parameters $v,w,\omega$. Because the
coordinates of the fermions enter Eq. (\ref{eq:S0}) only through the
differences $r_{j}\left(  \tau\right)  -r_{l}\left(  \sigma\right)  $, this
trial action is translation invariant.

Using the explicit forms of the exact (\ref{eq:Spol}) and the trial
(\ref{eq:S0}) actions, the variational inequality (\ref{VarIneq}) takes the
form%
\begin{align}
F\left(  \beta|N_{\uparrow},N_{\downarrow}\right)   &  \leq F_{0}\left(
\beta|N_{\uparrow},N_{\downarrow}\right)  +\frac{U}{2\beta}\int_{0}^{\beta
}\left\langle \sum_{j,l=1,\neq j}^{N}\frac{1}{\left\vert \mathbf{r}%
_{j}\mathbf{\left(  \tau\right)  -r}_{l}\mathbf{\left(  \tau\right)
}\right\vert }\right\rangle _{0}d\tau\nonumber\\
&  -\frac{\omega^{2}+w^{2}-v^{2}}{4N\beta}\int_{0}^{\beta}\left\langle
\sum_{j,l=1}^{N}\left(  \mathbf{r}_{j}\left(  \tau\right)  \mathbf{-r}%
_{l}\left(  \tau\right)  \right)  ^{2}\right\rangle _{0}d\tau\nonumber\\
&  -\frac{w}{8}\frac{v^{2}-w^{2}}{N\beta}\int_{0}^{\beta}\int_{0}^{\beta
}\left\langle \sum_{j,l=1}^{N}\left(  \mathbf{r}_{j}\left(  \tau\right)
-\mathbf{r}_{l}\left(  \sigma\right)  \right)  ^{2}\right\rangle _{0}%
\frac{\cosh w\left(  \frac{1}{2}\beta-\left\vert \tau-\sigma\right\vert
\right)  }{\sinh\frac{1}{2}\beta w}d\sigma d\tau\nonumber\\
&  -\frac{1}{2\beta}\int_{0}^{\beta}\int_{0}^{\beta}\sum_{\mathbf{k}%
}\left\vert V_{k}\right\vert ^{2}\left\langle \sum_{j,l=1}^{N}%
e^{i\mathbf{k\cdot}\left(  \mathbf{r}_{j}\left(  \tau\right)  -\mathbf{r}%
_{l}\left(  \sigma\right)  \right)  }\right\rangle _{0}\frac{\cosh
\omega_{\mathrm{LO}}\left(  \frac{1}{2}\beta-\left\vert \tau-\sigma\right\vert
\right)  }{\sinh\frac{1}{2}\beta\omega_{\mathrm{LO}}}d\sigma d\tau.
\label{VI1}%
\end{align}
and it is clear that the minimization automatically implies $v^{2}\geq w^{2}$.

In the zero-temperature limit ($\beta\rightarrow\infty$), we arrive after some
lengthy algebra at the following upper bound for the ground-state energy
$E^{0}\left(  N_{\uparrow},N_{\downarrow}\right)  $ of a translation invariant
$N$-polaron system%
\[
E^{0}\left(  N_{\uparrow},N_{\downarrow}\right)  \leq E_{var}\left(
N_{\uparrow},N_{\downarrow}|v,w,\omega\right)  ,
\]
with%
\begin{align}
E_{var}\left(  N_{\uparrow},N_{\downarrow}|v,w,\omega\right)   &  =\frac{3}%
{4}\frac{\left(  v-w\right)  ^{2}}{v}-\frac{3}{4}\omega+\frac{1}{2}%
\mathbb{E}_{F}\left(  N_{\downarrow}\right)  +\frac{1}{2}\mathbb{E}_{F}\left(
N_{\downarrow}\right) \nonumber\\
&  +E_{C\Vert}\left(  N_{\uparrow}\right)  +E_{C\Vert}\left(  N_{\downarrow
}\right)  +E_{C\uparrow\downarrow}\left(  N_{\uparrow},N_{\downarrow}\right)
\nonumber\\
&  +E_{\alpha\Vert}\left(  N_{\uparrow}\right)  +E_{\alpha\Vert}\left(
N_{\downarrow}\right)  +E_{\alpha\uparrow\downarrow}\left(  N_{\uparrow
},N_{\downarrow}\right)  , \label{Egr}%
\end{align}
where $\mathbb{E}_{F}\left(  N\right)  $ is the energy of $N$ spin-polarized
fermions confined to a parabolic potential with the confinement frequency
$\omega$, $E_{C\Vert}\left(  N_{\uparrow\left(  \downarrow\right)  }\right)  $
is the Coulomb energy of the electrons with parallel spins, $E_{C\uparrow
\downarrow}\left(  N_{\uparrow},N_{\downarrow}\right)  $ is the Coulomb energy
of the electrons with opposite spins, $E_{\alpha\Vert}\left(  N_{\uparrow
\left(  \downarrow\right)  }\right)  $ is the electron-phonon energy of the
electrons with parallel spins, and $E_{\alpha\uparrow\downarrow}\left(
N_{\uparrow},N_{\downarrow}\right)  $ is the electron-phonon energy of the
electrons with opposite spins. The key steps in the derivation and the
resulting analytical expressions for the terms of Eq. (\ref{Egr}) can be found
in the Appendix.

\section{Discussion of results}

In the present section we summarize and discuss the main results of the
numerical minimization of $E_{var}\left(  N_{\uparrow},N_{\downarrow
}|v,w,\omega\right)  $ with respect to the three variational parameters $v$,
$w$, and $\omega$. The Fr\"{o}hlich constant $\alpha$ and the Coulomb
parameter
\begin{equation}
\alpha_{0}\equiv\frac{U}{\sqrt{2}}\equiv\frac{\alpha}{1-\eta}\text{ with
}\frac{1}{\eta}=\frac{\varepsilon_{0}}{\varepsilon_{\infty}}%
\end{equation}
characterize the strength of the electron-phonon and of the Coulomb
interaction, obeying the physical condition $\alpha\geq\alpha_{0}$ [see
(\ref{units})]. The optimal values of the variational parameters $v,$$w,$and
$\omega$ are denoted $v_{op},$$w_{op},$and $\omega_{op}$, respectively. The
optimal value of the total spin was always determined by choosing the
combination $\left(  N_{\uparrow},N_{\downarrow}\right)  $ for fixed
$N=N_{\uparrow}+N_{\downarrow}$which corresponds to the lowest value
$E^{0}\left(  N\right)  $ of the variational functional%
\begin{equation}
E^{0}\left(  N\right)  \equiv\min_{N_{\uparrow}}E_{var}\left(  N_{\uparrow
},N-N_{\uparrow}|v_{op},w_{op},\omega_{op}\right)  .
\end{equation}
In Figs. 1 to 3, we present the ground-state energy per polaron (panel
\emph{a}), the addition energy (panel \emph{b}), the optimal values of the
variational parameters (panel \emph{c}) and the total spin (panel \emph{c}),
as a function of the number of polarons. The addition energy is determined by
the formula%
\begin{equation}
\Delta\left(  N\right)  \equiv E^{0}\left(  N+1\right)  -2E^{0}\left(
N\right)  +E^{0}\left(  N-1\right)  . \label{add}%
\end{equation}

In Fig. 1 we consider a highly polar system with $\alpha=\alpha_{0}=7$. The
optimal value $\omega_{op}$ (see panel \emph{c}) for the confinement frequency
$\omega$ is strictly positive (at least for $N\leq31)$. Therefore, the results
of Fig. 1 are related to multipolaron states analogous to those investigated
in Ref. \cite{mpol}. This interpretation is confirmed by the fact that (see
panel \emph{a} of Fig. 1) the ground-state energy per polaron for $N=2$ is
lower than that for $N=1$. For $N>2,$ the ground-state energy per polaron is
an increasing function of $N,$ which means that the effective electron-phonon
coupling weakens due to screening when the number of polarons increases.

The addition energy (panel \emph{b} of Fig. 1) oscillates, taking local maxima
at even $N$ and local minima at odd $N$. This oscillating behavior reflects
the trend of a stable multipolaron state to have the minimal possible spin.
This trend is an analog of the pairing of electrons in a superconducting
state. For even $N$ (see panel \emph{d} of Fig. 1) the total spin $S$ is equal
to zero. For odd $N,$ one electron remains non-paired and $S=1/2$. Therefore,
one intuitively expects that the states with $S=0$ are energetically favorable
as compared to states with $S=1/2,$ and hence, $\Delta\left(  N\right)  $ for
odd $N$ is lower than $\Delta\left(  N\right)  $ for even $N.$The plot of the
addition energies in panel \emph{b} of Fig. 1 confirms this expectation.
Furthermore, pronounced maxima in $\Delta\left(  N\right)  $ correspond to
closed-shell systems with $N=2,8,20...$.

The optimal values of the variational parameters (panel \emph{c} of Fig. 1)
reveal a general trend to decrease as a function of $N$, except the parameter
$v$, which have a peak at $N=2$. This peak, as well as the minimum of
$E^{0}\left(  N\right)  /N$ at $N=2,$shows that the two-polaron state in the
extremely strong-coupling regime is especially stable with respect to the
other multipolaron states with $N>2$. The dependence of the parameter $\omega$
on $N$ starts from $N=2,$ because the one-polaron variational functional does
not depend on $\omega$.

In Fig. 2, the ground state energy, the additional energy, the variational
parameters, and the total spin for $N$-polaron systems are plotted for
$\alpha=3$, $\alpha_{0}=4.5$, and $\eta=1/3$. In this regime, the optimal
value for the confinement frequency $\omega$ is $\omega_{op}=0$ (panel
\emph{c}). Therefore, in this regime, as well as at weaker electron-phonon
coupling strengths, $N>1$ polarons do not form a multipolaron state. The
addition energy, as seen from panel \emph{b} of Fig. 2, has no oscillations or
peaks in the case when $N$ polarons do not form a multipolaron state. It
should be noted, that in the case when $\omega_{op}=0$, we deal with a finite
number $N$ of polarons in an infinite volume. So, at $\omega_{op}=0$ the
many-body effects, related to the electron-electron interaction and to the
Fermi statistics, are vanishingly small. The dependence of the ground-state
energy of the total spin of a many-polaron system is just one of these
many-body effects. As a consequence, the ground-state energy within the
present model at $\omega_{op}=0$ does not depend on the total spin. For this
reason, there is no panel \emph{d} in Fig. 2.

Figs. 1 and 2 represent two mutually opposite cases (with $\omega\neq0$ and
with $\omega_{op}=0$ for all $N$ under consideration). Fig. 3 describes the
case when the regime with $\omega_{op}\neq0$ (for $N\leq16$) changes to the
regime with $\omega_{op}=0$ (for $N\geq17$). As seen from panel \emph{a} of
Fig. 3, the ground-state energy for $N\leq16$ behaves similarly to that
calculated for $\alpha=\alpha_{0}=7$ (panel \emph{a} of Fig. 1), with the
following distinction: for $\alpha_{0}/\alpha=1.01$ ($\alpha=7$) it appears
that$\left.  E^{0}\left(  N\right)  /N\right\vert _{N=2}>\left.  E^{0}\left(
N\right)  /N\right\vert _{N=1}$, while for $\alpha_{0}/\alpha=1$ ($\alpha=7$),
$\left.  E^{0}\left(  N\right)  /N\right\vert _{N=2}<\left.  E^{0}\left(
N\right)  /N\right\vert _{N=1}$. As seen from panel \emph{c} of Fig. 3, when
an extra polaron is added to $N=16$ polarons, the optimal value for $\omega$
switches to zero, and therefore, the multipolaron state transforms to the
ground state of $N$ independent polarons. When $N$ changes from $N=16$ to
$N=17$, the ground-state energy per polaron slightly jumps down and is
practically constant with further increasing $N.$ The transition from a
multipolaron state to a state of $N$ independent polarons is clearly revealed
in the dependence of the addition energy on the number of polarons (panel
\emph{b} of Fig. 3). At $N=16,$ $\Delta\left(  N\right)  $ has a pronounced
minimum, which is a manifestation of the transition from a multipolaron ground
state to a ground state of $N$ independent polarons.

The total spin, as seen from panel \emph{d} of Fig. 3, takes its minimal
possible value for $N\leq13,$ while for $N\geq14$, the ground state is
spin-polarized. So, the transition from the ground state with the minimal
possible spin to the spin-polarized ground state with increasing $N$ precedes
the break-up of a multipolaron state. For $N\geq17,$ in the same way as in the
case $\left(  \alpha=3,\alpha_{0}=4.5\right)  ,$ the variational ground-state
energy of an $N$-polaron system does not depend on the total spin.

In Fig. 4, the \textquotedblleft phase diagrams\textquotedblright\ analogous
to that of Ref. \cite{VPD91} are plotted for an $N$-polaron system in bulk
with $N=2,3,5,$ and $10$. The area where $\alpha_{0}\leq\alpha$ is the
non-physical region. For $\alpha>\alpha_{0}$, each sector between a curve
corresponding to a well defined $N$ and the line indicating $\alpha_{0}%
=\alpha$ shows the stability region where $\omega_{op}\neq0$, while the white
area corresponds to the regime with $\omega_{op}=0$. When comparing the
stability region for $N=2$ from Fig. 4 with the bipolaron \textquotedblleft
phase diagram\textquotedblright\ of Ref. \cite{VPD91}, the stability region in
the present work starts from the value $\alpha_{c}\approx4.1$ (instead of
$\alpha_{c}\approx6.9$ in Ref. \cite{VPD91}). The width of the stability
region within the present model is also larger than the width of the stability
region within the model of Ref. \cite{VPD91}. Also, the absolute values of the
ground-state energy of a two-polaron system given by the present model are
smaller than those given by the approach of Ref. \cite{VPD91}.

The difference between the numerical results of the present work and of Ref.
\cite{VPD91} is due to the following distinction between the used model
systems. The model system of Ref. \cite{VPD91} consists of two electrons
interacting with two fictitious particles and with each other through
quadratic interactions. But the trial Hamiltonian given by Eq. (6) of Ref.
\cite{VPD91} is not symmetric with respect to the permutation of the
electrons. It is only symmetric under the permutation of the pairs
\textquotedblleft electron + fictitious particle\textquotedblright. As a
consequence, this trial system is only applicable if the electrons are
distinguishable, i.e. have opposite spin. In contrast to the model of Ref.
\cite{VPD91}, the model used in the present paper is described by the trial
action (9), which is fully symmetric with respect to the permutations of the
electrons, as is required to describe identical particles. Up to now we have
been unable to construct such a model with two retardation sources. As a
consequence, the trial model of Ref. [27] is superior to our model for
describing 2 polarons because it has more varaiational parameters, but its
applicability is limited to 2 polarons. The generalization of the model of
Ref. \cite{VPD91} to $N>2$ is currently under investigation.

The \textquotedblleft phase diagrams\textquotedblright\ for $N>2$ demonstrate
the existence of stable multipolaron states (see Ref. \cite{mpol}). As
distinct from Ref. \cite{mpol}, here the ground state of an $N$-polaron system
is investigated supposing that the electrons are fermions. As seen from these
figures, for $N>2$, the stability region for a multipolaron state is narrower
than the stability region for $N=2$, and its width decreases with increasing
$N$. The critical value $\alpha_{c}$ for the electron-phonon coupling constant
increases with increasing $N$. From this behavior we can deduce a general
trend, which explains the behavior of the ground-state energy and related
quantities as a function of $N$ shown in Fig. 3. For fixed values of $\alpha$
and $\eta$, the width of the stability region for a multipolaron state is a
decreasing function of the number of electrons. Therefore, for any $\left(
\alpha,\eta\right)  $ there exists a critical number of electrons
$N_{0}\left(  \alpha,\eta\right)  $ such that a multipolaron state exists for
$N\leq N_{0}\left(  \alpha,\eta\right)  $ and does not exist for
$N>N_{0}\left(  \alpha,\eta\right)  $. For example, for the $N$-polaron system
described in Fig. 1, $N_{0}$ is at least larger than 20. For the system shown
in Fig. 2, $N_{0}=1$, and for the system in Fig.3, $N_{0}=16$. If we add
electrons to an $N$-polaron system one by one, the multipolaron state breaks
up when the number of electrons exceeds a critical value $N_{0}\left(
\alpha,\eta\right)  $.

In order to analyze the consequences of the Fermi statistics for the
ground-state properties of an $N$-polaron system, we compare the ground-state
energies calculated with and without the Fermi statistics. In Table 1, the
results are presented for the ground-state energy per particle in units of the
one-polaron strong-coupling energy $E_{1}$,%
\begin{equation}
\mathcal{E}_{N}=\frac{E^{0}\left(  N\right)  }{NE_{1}}\;\;\left(  E_{1}%
\equiv\frac{1}{3\pi}\alpha^{2}\right)
\end{equation}
with $\alpha=10$, $\eta=0$ for three cases: the many-body path-integral
approach of the present work with fermion statistics ($\mathcal{E}%
_{N}^{\left(  F\right)  }$), the same approach for distinguishable particles
($\mathcal{E}_{N}^{\left(  d\right)  }$), and the strong-coupling approach of
Ref. \cite{mpol}, which also does not take into account the Fermi statistics
($\mathcal{E}_{N}^{\left(  d,sc\right)  }$).

\bigskip

\begin{center}
Table 1. The polaron characteristic energy $\mathcal{E}_{N}$ calculated using
different methods.

\medskip%

\begin{tabular}
[c]{|l|l|l|l|}\hline\hline
\multicolumn{1}{||l|}{$N$} & \multicolumn{1}{||l|}{$\mathcal{E}_{N}^{\left(
F\right)  }$} & \multicolumn{1}{||l|}{$\mathcal{E}_{N}^{\left(  d\right)  }$}
& \multicolumn{1}{||l||}{$\mathcal{E}_{N}^{\left(  d,sc\right)  }$%
}\\\hline\hline
$2$ & $-1.349$ & $-1.349$ & $-1.148$\\\hline
$3$ & $-1.308$ & $-1.415$ & $-1.241$\\\hline
$4$ & $-1.296$ & $-1.468$ & $-1.308$\\\hline
$5$ & $-1.279$ & $-1.508$ & $-1.361$\\\hline
$6$ & $-1.272$ & $-1.536$ & $-1.404$\\\hline
\end{tabular}

\end{center}

\medskip

As seen from Table 1, the ground-state energy per particle for $N$ identical
polarons $\mathcal{E}_{N}^{\left(  F\right)  }$ is higher than that for $N$
distinguishable polarons $\mathcal{E}_{N}^{\left(  d\right)  }$. Furthermore,
$\mathcal{E}_{N}^{\left(  F\right)  }$ increases whereas $\mathcal{E}%
_{N}^{\left(  d\right)  }$ decreases with an increasing number of polarons.
Note however that $\mathcal{E}_{N}^{\left(  d\right)  }<\mathcal{E}%
_{N}^{\left(  d,sc\right)  }$ for the considered values of $\alpha$ and $\eta
$, which means that the path-integral variational method provides better
results for the $N$-polaron ground-state energy than the strong-coupling
approach \cite{mpol} (at least for $\alpha\leq10$).

Another consequence of the Fermi statistics is the dependence of the polaron
characteristics and of the total spin of an $N$-polaron system on the
parameters ($\alpha,\,\alpha_{0},$$N$). In Fig. 5, we present the ground-state
energy per particle, the confinement frequency $\omega_{op}$ and the total
spin $S$ as a function of the coupling constant $\alpha$ for $\alpha
_{0}/\alpha=1.05$ and for a different numbers of polarons. The ground-state
energy turns out to be a continuous function of $\alpha$, while $\omega_{op}%
$and $S$ reveal jumps. For all considered numbers of polarons $N>2$, there is
a region of $\alpha$ in which $S$ takes its maximal value, while $\omega
_{op}\neq0$. When lowering $\alpha$, this spin-polarized state with parallel
spins precedes the transition from the regime with $\omega_{op}\neq0$ to the
regime with $\omega_{op}=0$ (the break-up of a multipolaron state). For $N=2$
(the case of a bipolaron), we see from Fig. 5 that the ground state has a
total spin $S=0$ for all values of $\alpha$, i.~e., the ground state of a
bipolaron is a singlet. This result is in agreement with earlier
investigations on the large-bipolaron problem (see, e.~g.,
\cite{Kashirina2003}).

\section{Conclusions}

Using the extension of the Jensen-Feynman variational principle to the systems
of identical particles, we have derived a rigorous upper bound for the free
energy of a translation invariant system of $N$ interacting polarons. In the
zero-temperature limit, the variational free energy provides the variational
functional for the ground-state energy of the $N$-polaron system. The
developed approach is valid for an arbitrary coupling strength. The resulting
ground-state energy is obtained taking into account the Fermi statistics of electrons.

For sufficiently high values of the electron-phonon coupling constant and of
the ratio $1/\eta=\varepsilon_{0}/\varepsilon_{\infty}$, the system of $N$
interacting polarons can form a stable multipolaron ground state. When this
state is formed, the total spin of the system takes its minimal possible
value. The larger the number of electrons, the narrower the stability region
of a multipolaron state becomes. So, when adding electrons one by one to a
stable multipolaron state, it breaks up for a definite number of electrons
$N_{0}$, which depends on the coupling constant and on the ratio of the
dielectric constants. This break-up is preceded by the change from a
spin-mixed ground state with a minimal possible spin to a spin-polarized
ground state with parallel spins.

For a stable multipolaron state, the addition energy reveals peaks
corresponding to closed shells. At $N=N_{0}$, the addition energy has a
pronounced minimum. These features of the addition energy, as well as the
total spin as a function of the number of electrons, might be resolved
experimentally using, e.g., capacity and magnetization measurements.

\begin{acknowledgments}
This work has been supported by the GOA BOF UA 2000, IUAP, FWO-V projects
G.0306.00, G.0274.01N, G.0435.03, the WOG WO.025.99N (Belgium) and the
European Commission GROWTH Programme, NANOMAT project, contract No. G5RD-CT-2001-00545.
\end{acknowledgments}

\appendix

\section{Mathematical details}

\subsection{Generalization of the Hellman-Feynman theorem}%

\def\theequation{A.\arabic{equation}}
\setcounter{equation}{0}%

For the averages of the quadratic terms in Eq. (\ref{VI1}), we can derive a
generalization of the Hellman-Feynman theorem for the case where we have a
(trial) action but no Hamiltonian. Indeed, since $F_{0}=-\frac{1}{\beta}\ln
Z_{0}$ it follows that
\begin{equation}
\frac{d}{d\gamma}F_{0}=-\frac{1}{\beta}\frac{d}{d\gamma}\ln Z_{0}=-\frac
{1}{\beta}\frac{1}{Z_{0}}\frac{d}{d\gamma}Z_{0}=-\frac{1}{\beta}\left\langle
\frac{dS_{0}}{d\gamma}\right\rangle _{0} \label{eq:Hellman-Feynman}%
\end{equation}
for any parameter $\gamma$ in the trial action. Taking the derivative of
$S_{0}$ [eq. (\ref{eq:S0})] with respect to $\omega$ and $v$ then gives%
\begin{align*}
\int_{0}^{\beta}\left\langle \sum_{j,l=1}^{N}\left(  \mathbf{r}_{j}\left(
\tau\right)  \mathbf{-r}_{l}\left(  \tau\right)  \right)  ^{2}\right\rangle
_{0}d\tau &  =-\frac{2N}{\omega}\left\langle \frac{dS_{0}}{d\omega
}\right\rangle _{0}=\frac{2N\beta}{\omega}\frac{dF_{0}}{d\omega},\\
\sum_{j,l=1}^{N}\int_{0}^{\beta}\int_{0}^{\beta}\left\langle \sum_{j,l=1}%
^{N}\left(  \mathbf{r}_{j}\left(  \tau\right)  -\mathbf{r}_{l}\left(
\sigma\right)  \right)  ^{2}\right\rangle _{0}\frac{\cosh w\left(  \frac{1}%
{2}\beta-\left\vert \tau-\sigma\right\vert \right)  }{\sinh\frac{1}{2}\beta
w}d\sigma d\tau &  =\frac{4N\beta}{wv}\left(  \frac{dF_{0}}{dv}+\frac
{v}{\omega}\frac{dF_{0}}{d\omega}\right)  ,
\end{align*}
and therefore the variational inequality becomes%
\begin{align}
F\left(  \beta|N_{\uparrow},N_{\downarrow}\right)   &  \leq F_{0}\left(
\beta|N_{\uparrow},N_{\downarrow}\right)  -\frac{1}{2}\omega\frac
{dF_{0}\left(  \beta|N_{\uparrow},N_{\downarrow}\right)  }{d\omega}-\frac
{1}{2}\frac{v^{2}-w^{2}}{v}\frac{dF_{0}\left(  \beta|N_{\uparrow
},N_{\downarrow}\right)  }{dv}\nonumber\\
&  +\frac{U}{2\beta}\int_{0}^{\beta}\left\langle \sum_{j,l=1,\neq j}^{N}%
\frac{1}{\left\vert \mathbf{r}_{j}\mathbf{\left(  \tau\right)  -r}%
_{l}\mathbf{\left(  \tau\right)  }\right\vert }\right\rangle _{0}%
d\tau\nonumber\\
&  -\frac{1}{2\beta}\int_{0}^{\beta}\int_{0}^{\beta}\sum_{\mathbf{k}%
}\left\vert V_{k}\right\vert ^{2}\left\langle \sum_{j,l=1}^{N}%
e^{i\mathbf{k\cdot}\left(  \mathbf{r}_{j}\left(  \tau\right)  -\mathbf{r}%
_{l}\left(  \sigma\right)  \right)  }\right\rangle _{0}\frac{\cosh
\omega_{\mathrm{LO}}\left(  \frac{1}{2}\beta-\left\vert \tau-\sigma\right\vert
\right)  }{\sinh\frac{1}{2}\beta\omega_{\mathrm{LO}}}d\sigma d\tau.
\label{VIn2}%
\end{align}

\subsection{Correlation and density functions}

In order to calculate the Coulomb and the electron-phonon energies [the terms
in the second and third lines of Eq. (\ref{VIn2}), respectively], we only need
the pair correlation function $g_{F}$ and the two-point correlation function
$C_{F}$ for fermions which we define as%
\begin{align}
g_{F}\left(  r,\beta|N_{\uparrow},N_{\downarrow}\right)   &  =\frac
{1}{N\left(  N-1\right)  }\sum_{j,l=1;j\neq l}^{N}\left\langle \delta\left(
\mathbf{r-r}_{j}+\mathbf{r}_{l}\right)  \right\rangle _{0},\label{gf}\\
C_{F}\left(  \mathbf{q,}\tau,\beta|N_{\uparrow},N_{\downarrow}\right)   &
=\frac{1}{N^{2}}\sum_{j,l=1}^{N}\left\langle e^{-i\mathbf{q\cdot r}_{l}\left(
\tau\right)  }e^{i\mathbf{q\cdot r}_{j}\left(  0\right)  }\right\rangle _{0},
\end{align}
where $\left\langle \ldots\right\rangle _{0}$ denotes a path-integral average
with the action functional $S_{0}$. After a separation of the center-of-mass
motion (see Ref. \cite{MPQD2004}), these correlation functions take the form%
\begin{align}
g_{F}\left(  r,\beta|N_{\uparrow},N_{\downarrow}\right)   &  =\frac{1}{N}%
\frac{1}{N-1}\sum_{j,l=1;j\neq l}^{N}\left\langle \delta\left(  \mathbf{r-r}%
_{j}+\mathbf{r}_{l}\right)  \right\rangle _{F},\label{GF}\\
C_{F}\left(  q\mathbf{,}\tau,\beta|N_{\uparrow},N_{\downarrow}\right)   &
=\mathbb{C}_{F}\left(  q\mathbf{,}\tau,\beta|N_{\uparrow},N_{\downarrow
}\right)  \exp\left[  -\frac{q^{2}}{N}\left(  \frac{w^{2}\tau\left(
\beta-\tau\right)  }{2v^{2}\beta}\right.  \right. \nonumber\\
&  +\frac{v^{2}-w^{2}}{v^{3}}\frac{\sinh\left(  \frac{1}{2}v\tau\right)
\sinh\left(  \frac{1}{2}v\left(  \beta-\tau\right)  \right)  }{\sinh\left(
\frac{1}{2}v\beta\right)  }\nonumber\\
&  \left.  \left.  -\frac{\sinh\left(  \frac{1}{2}\omega\tau\right)
\sinh\left(  \frac{1}{2}\omega\left(  \beta-\tau\right)  \right)  }%
{\omega\sinh\left(  \frac{1}{2}\omega\beta\right)  }\right)  \right]
\label{CF}%
\end{align}
with%
\begin{equation}
\mathbb{C}_{F}\left(  \mathbf{q,}\tau,\beta|N_{\uparrow},N_{\downarrow
}\right)  =\frac{1}{N^{2}}\sum_{j,l=1}^{N}\left\langle e^{-i\mathbf{q\cdot
r}_{l}\left(  \tau\right)  }e^{i\mathbf{q\cdot r}_{j}\left(  0\right)
}\right\rangle _{F}, \label{CFT}%
\end{equation}
where $\left\langle \ldots\right\rangle _{F}$ denotes a path-integral average
with the action functional%
\begin{equation}
S_{F}=-\frac{1}{2}\int_{0}^{\beta}\sum_{j=1}^{N}\left[  \left(  \frac
{d\mathbf{r}_{j}\left(  \tau\right)  }{d\tau}\right)  ^{2}+\omega
^{2}\mathbf{r}_{j}^{2}\left(  \tau\right)  \right]  d\tau\label{SF}%
\end{equation}
for $N=N_{\uparrow}+N_{\downarrow}$ independent fermions in a 3D parabolic
potential with the confinement frequency $\omega$. We shall use also the
density function%
\begin{equation}
\tilde{n}_{F}\left(  \mathbf{q,}\beta|N_{\uparrow},N_{\downarrow}\right)
=\frac{1}{N}\sum_{j=1}^{N}\left\langle e^{-i\mathbf{q\cdot r}_{j}%
}\right\rangle _{F}. \label{DF}%
\end{equation}
The functions (\ref{GF}), (\ref{CFT}), and (\ref{DF}) were already derived
before (see Refs. \cite{MPQD2004,CF1,CF2}).

Both the Coulomb energy and the electron-phonon energy in Eq. (\ref{VIn2}) are
effectively Coulomb terms but with two important differences. Firstly, the
standard Coulomb repulsion between the electrons is static, whereas the
effective Coulomb attraction due to the polaron effect is retarded. A direct
consequence of this difference is that the center of mass plays no role in the
Coulomb repulsion, whereas it is essential in the retarded contribution.
Secondly, the self-interaction has to be excluded from the Coulomb repulsion,
whereas it contributes in the electron-phonon contribution. This is the main
reason why we treat the Coulomb repulsion via the pair correlation function
(in real space), and the retarded interaction with the two-point correlation
function $C_{F}\left(  \mathbf{k,}\tau,\beta|N\right)  $ (i.e. in wave number
space). In principle we have the choice to handle both terms either in real
space or in wave number space.

Having the definitions of the pair correlation function and the two point
correlation function in mind, we thus obtain for the free energy (performing
the angular integrations at once)%
\begin{align}
F\left(  \beta|N_{\uparrow},N_{\downarrow}\right)   &  \leq F_{0}\left(
\beta|N_{\uparrow},N_{\downarrow}\right)  -\frac{1}{2}\omega\frac
{dF_{0}\left(  \beta|N_{\uparrow},N_{\downarrow}\right)  }{d\omega}-\frac
{1}{2}\frac{v^{2}-w^{2}}{v}\frac{dF_{0}\left(  \beta|N_{\uparrow
},N_{\downarrow}\right)  }{dv}\nonumber\\
&  +2\pi U\int_{0}^{\infty}rN\left(  N-1\right)  g_{F}\left(  r,\beta
|N_{\uparrow},N_{\downarrow}\right)  dr\nonumber\\
&  -\frac{\sqrt{2}\alpha}{\pi}\int_{0}^{\beta/2}\int_{0}^{\infty}N^{2}%
C_{F}\left(  q\mathbf{,}\tau,\beta|N_{\uparrow},N_{\downarrow}\right)
dq\frac{\cosh\left(  \frac{1}{2}\beta-\tau\right)  }{\sinh\left(  \frac{1}%
{2}\beta\right)  }d\tau. \label{VIn3}%
\end{align}
For the ground state energy $\left(  \beta\rightarrow\infty\right)  $ we thus
find%
\[
E^{0}\left(  N_{\uparrow},N_{\downarrow}\right)  \leq E_{var}\left(
N_{\uparrow},N_{\downarrow}|v,w,\omega\right)  ,
\]
where the variational functional is%
\begin{align}
E_{var}\left(  N_{\uparrow},N_{\downarrow}|v,w,\omega\right)   &
=E_{0}\left(  N_{\uparrow},N_{\downarrow}\right)  -\frac{1}{2}\omega
\frac{dE_{0}\left(  N_{\uparrow},N_{\downarrow}\right)  }{d\omega}-\frac{1}%
{2}\frac{v^{2}-w^{2}}{v}\frac{dE_{0}\left(  N_{\uparrow},N_{\downarrow
}\right)  }{dv}\nonumber\\
&  +2\pi U\int_{0}^{\infty}rN\left(  N-1\right)  g_{F}\left(  r,\beta
\rightarrow\infty|N_{\uparrow},N_{\downarrow}\right)  dr\nonumber\\
&  -\frac{\sqrt{2}\alpha}{\pi}\int_{0}^{\infty}\int_{0}^{\infty}N^{2}%
C_{F}\left(  q\mathbf{,}\tau,\beta\rightarrow\infty|N_{\uparrow}%
,N_{\downarrow}\right)  dqe^{-\tau}d\tau. \label{Eg}%
\end{align}
Here, $E_{0}$ is the ground state energy corresponding to the trial action,
given by%
\begin{equation}
E_{0}\left(  N_{\uparrow},N_{\downarrow}\right)  =\frac{3}{2}\left(
v-w-\omega\right)  +\mathbb{E}_{F}\left(  N_{\uparrow}\right)  +\mathbb{E}%
_{F}\left(  N_{\downarrow}\right)  , \label{E0}%
\end{equation}
where $\mathbb{E}_{F}\left(  N\right)  $ is the ground state energy of $N$
fermions with parallel spins and with energy levels $\epsilon_{n}=\left(
n+\frac{3}{2}\right)  \omega$:
\begin{equation}
\mathbb{E}_{F}\left(  N\right)  =\frac{1}{8}L\left(  L+2\right)  \left(
L+1\right)  ^{2}\omega+\left(  N-N_{L}\right)  \left(  L+\frac{3}{2}\right)
\omega, \label{EF}%
\end{equation}
$L$ denotes the lowest partially filled or empty level, and
\[
N_{L}=\frac{1}{6}L\left(  L+1\right)  \left(  L+2\right)
\]
is the number of fermions in the fully occupied levels. Filling out $E_{0}$ in
$E_{var}$ we thus obtain%
\begin{align}
E_{var}\left(  N_{\uparrow},N_{\downarrow}|v,w,\omega\right)   &  =\frac{3}%
{4}\frac{\left(  v-w\right)  ^{2}}{v}-\frac{3}{4}\omega+\frac{1}{2}%
\mathbb{E}_{F}\left(  N_{_{\uparrow}}\right)  +\frac{1}{2}\mathbb{E}%
_{F}\left(  N_{\downarrow}\right) \nonumber\\
&  +2\pi U\int_{0}^{\infty}rN\left(  N-1\right)  g_{F}\left(  r,\beta
\rightarrow\infty|N_{\uparrow},N_{\downarrow}\right)  dr\nonumber\\
&  -\frac{\sqrt{2}\alpha}{\pi}\int_{0}^{\infty}\int_{0}^{\infty}N^{2}%
C_{F}\left(  q\mathbf{,}\tau,\beta\rightarrow\infty|N_{\uparrow}%
,N_{\downarrow}\right)  dqe^{-\tau}d\tau, \label{Eg2}%
\end{align}
where the factor 1/2 in front of $\mathbb{E}_{F}$ is a consequence of the
subtraction $\mathbb{E}_{F}-\frac{1}{2}\omega\frac{d\mathbb{E}_{F}}{d\omega},$
not surprising because of the virial theorem for harmonic oscillators. The
term $\frac{3}{4}\frac{\left(  v-w\right)  ^{2}}{v}$ is precisely the same as
in Feynman's treatment of the polaron, but of course the values of $v$ and $w$
will be different if many-particle effects will be taken into account.

We now split $g_{F}$ and $C_{F}$ for a mixture of fermions with different spin
projections into the parts corresponding to parallel and opposite spins. The
case of $N_{\uparrow}$ electrons with spin up and $N_{\downarrow}$ electrons
with spin down can be found after some reflection in terms of the
spin-polarized quantities:%
\begin{align}
N\tilde{n}_{F}\left(  \mathbf{q,}\beta|N=N_{\uparrow}+N_{\downarrow}\right)
&  =N_{\uparrow}\tilde{n}_{F}\left(  \mathbf{q,}\beta|N_{\uparrow}\right)
+N_{\downarrow}\tilde{n}_{F}\left(  \mathbf{q,}\beta|N_{\downarrow}\right)
,\label{eq:mixture1}\\
N\left(  N-1\right)  g_{F}\left(  r,\beta|N_{\uparrow},N_{\downarrow}\right)
&  =N_{\uparrow}\left(  N_{\uparrow}-1\right)  g_{F}\left(  r,\beta
|N_{\uparrow}\right)  +N_{\downarrow}\left(  N_{\downarrow}-1\right)
g_{F}\left(  r,\beta|N_{\downarrow}\right) \nonumber\\
&  +\frac{2N_{\uparrow}N_{\downarrow}}{\left(  2\pi\right)  ^{3}}\int
e^{i\mathbf{q\cdot r}}\tilde{n}_{F}\left(  \mathbf{q,}\beta|N_{\uparrow
}\right)  \tilde{n}_{F}\left(  \mathbf{q,}\beta|N_{\downarrow}\right)
d\mathbf{q,}\label{eq:mixture2}\\
N^{2}\mathbb{C}_{F}\left(  \mathbf{q,}\tau,\beta|N_{\uparrow},N_{\downarrow
}\right)   &  =N_{\uparrow}^{2}\mathbb{C}_{F}\left(  \mathbf{q,}\tau
,\beta|N_{\uparrow}\right)  +N_{\downarrow}^{2}\mathbb{C}_{F}\left(
\mathbf{q,}\tau,\beta|N_{\downarrow}\right) \nonumber\\
&  +2N_{\uparrow}N_{\downarrow}\tilde{n}_{F}\left(  \mathbf{q,}\beta
|N_{\uparrow}\right)  \tilde{n}_{F}\left(  \mathbf{q,}\beta|N_{\downarrow
}\right)  . \label{eq:mixture3}%
\end{align}

\subsection{Coulomb and electron-phonon energies}

Using Eqs. (\ref{eq:mixture1}) to (\ref{eq:mixture3}) in the variational
functional (\ref{Eg}), we arrive at the expression with three Coulomb terms
and three electron-phonon terms as follows:%
\begin{align}
E_{var}\left(  N_{\uparrow},N_{\downarrow}|v,w,\omega\right)   &  =\frac{3}%
{4}\frac{\left(  v-w\right)  ^{2}}{v}-\frac{3}{4}\omega+\frac{1}{2}%
\mathbb{E}_{F}\left(  N_{\downarrow}\right)  +\frac{1}{2}\mathbb{E}_{F}\left(
N_{\downarrow}\right) \nonumber\\
&  +E_{C\Vert}\left(  N_{\uparrow}\right)  +E_{C\Vert}\left(  N_{\downarrow
}\right)  +E_{C\uparrow\downarrow}\left(  N_{\uparrow},N_{\downarrow}\right)
\nonumber\\
&  +E_{\alpha\Vert}\left(  N_{\uparrow}\right)  +E_{\alpha\Vert}\left(
N_{\downarrow}\right)  +E_{\alpha\uparrow\downarrow}\left(  N_{\uparrow
},N_{\downarrow}\right)  \label{Eg4}%
\end{align}
with the contributions%
\begin{align}
E_{C\Vert}\left(  N\right)   &  =2\pi N\left(  N-1\right)  U\int_{0}^{\infty
}rg_{F}\left(  r,\beta\rightarrow\infty|N\right)  dr,\label{t0}\\
E_{C\uparrow\downarrow}\left(  N_{\uparrow},N_{\downarrow}\right)   &  =4\pi
N_{\uparrow}N_{\downarrow}U\int_{0}^{\infty}r\frac{1}{\left(  2\pi\right)
^{3}}\int e^{i\mathbf{q\cdot r}}\tilde{n}_{F}\left(  \mathbf{q,}%
\beta\rightarrow\infty|N_{\uparrow}\right)  \tilde{n}_{F}\left(
\mathbf{q,}\beta\rightarrow\infty|N_{\downarrow}\right)  d\mathbf{q}%
dr,\label{t1}\\
E_{\alpha\Vert}\left(  N\right)   &  =-\frac{\sqrt{2}\alpha}{\pi}N^{2}\int
_{0}^{\infty}\int_{0}^{\infty}e^{-\frac{q^{2}}{2\left(  N_{\uparrow
}+N_{\downarrow}\right)  }\left(  \frac{w^{2}}{v^{2}}\tau+\frac{v^{2}-w^{2}%
}{v^{3}}\left(  1-e^{-v\tau}\right)  -\frac{1-e^{-\omega\tau}}{\omega}\right)
-\tau}\mathbb{C}_{F}\left(  \mathbf{q,}\tau,\beta\rightarrow\infty|N\right)
dqd\tau,\label{t2}\\
E_{\alpha\uparrow\downarrow}\left(  N_{\uparrow},N_{\downarrow}\right)   &
=-\frac{2\sqrt{2}\alpha}{\pi}N_{\uparrow}N_{\downarrow}\int_{0}^{\infty}%
dq\int_{0}^{\infty}d\tau e^{-\frac{q^{2}}{2\left(  N_{\uparrow}+N_{\downarrow
}\right)  }\left(  \frac{w^{2}}{v^{2}}\tau+\frac{v^{2}-w^{2}}{v^{3}}\left(
1-e^{-v\tau}\right)  -\frac{1-e^{-\omega\tau}}{\omega}\right)  -\tau
}\nonumber\\
&  \times\tilde{n}_{F}\left(  \mathbf{q,}\beta\rightarrow\infty|N_{\uparrow
}\right)  \tilde{n}_{F}\left(  \mathbf{q,}\beta\rightarrow\infty
|N_{\downarrow}\right)  . \label{t3}%
\end{align}
The integrations over $\mathbf{q}$ and over $r$ in Eqs. (\ref{t0}) to
(\ref{t3}) are performed using the explicit form of the density and
correlation functions (see Eq. \cite{MPQD2004})%
\begin{align}
\tilde{n}_{F}\left(  \mathbf{q,}\beta\rightarrow\infty|N\right)   &  =\frac
{1}{N}\sum_{k=0}^{L}n_{k}\left(  q\right)  f_{1}\left(  k|\beta\rightarrow
\infty,N\right)  ,\label{nf}\\
\mathbb{C}_{F}\left(  \mathbf{q,}\tau,\beta\rightarrow\infty|N\right)   &
=\frac{1}{N^{2}}\sum_{k=0}^{L}\sum_{k^{\prime}=L}^{\infty}M_{kk^{\prime}%
}\left(  q\right)  e^{\left(  k-k^{\prime}\right)  \omega\tau}\left[
f_{1}\left(  k|\beta\rightarrow\infty,N\right)  -f_{2}\left(  k,k^{\prime
}|\beta\rightarrow\infty,N\right)  \right]  ,\label{ccf}\\
g_{F}\left(  r,\beta\rightarrow\infty|N\right)   &  =\frac{1}{N\left(
N-1\right)  }\frac{1}{\left(  2\pi\right)  ^{3}}\int d\mathbf{q}%
e^{i\mathbf{q\cdot r}}\sum_{k=0}^{L}\sum_{k^{\prime}=L}^{\infty}M_{kk^{\prime
}}\left(  q\right) \nonumber\\
&  \times\left[  f_{1}\left(  k|\beta\rightarrow\infty,N\right)  -f_{2}\left(
k,k^{\prime}|\beta\rightarrow\infty,N\right)  \right]  \label{ggf}%
\end{align}
with the matrix elements
\begin{align}
n_{k}\left(  q\right)   &  =\exp\left(  -\frac{q^{2}}{4\omega}\right)
L_{k}^{\left(  2\right)  }\left(  \frac{q^{2}}{2\omega}\right)  ,\label{fn}\\
M_{kk^{\prime}}\left(  q\right)   &  =e^{-\frac{q^{2}}{2\omega}}\left(
\frac{q^{2}}{2\omega}\right)  ^{k_{>}-k_{<}}\sum_{j=0}^{k_{<}}\left(
j+1\right)  \frac{\left(  k_{<}-j\right)  !}{\left(  k_{>}-j\right)  !}\left[
L_{k_{<}-j}^{\left(  k_{>}-k_{<}\right)  }\left(  \frac{q^{2}}{2\omega
}\right)  \right]  ^{2}\quad\left(
\begin{array}
[c]{c}%
k_{<}\equiv\min\left(  k,k^{\prime}\right)  ,\\
k_{>}\equiv\max\left(  k,k^{\prime}\right)
\end{array}
\right)  , \label{bnn}%
\end{align}
where $L_{k}^{\left(  \alpha\right)  }\left(  x\right)  $ are the Laguerre
polynomials, and with one-particle and two-particle distribution functions%
\begin{equation}
f_{1}\left(  k|\beta\rightarrow\infty,N\right)  =\left\{
\begin{array}
[c]{cc}%
1, & k<L,\\
0, & k>L,\\
\frac{N-N_{L}}{N_{L+1}-N_{L}}, & k=L,
\end{array}
\right.  \label{p1}%
\end{equation}%
\begin{equation}
f_{2}\left(  k,k^{\prime}|\beta\rightarrow\infty,N\right)  =\left\{
\begin{array}
[c]{cc}%
f_{1}\left(  k|\beta\rightarrow\infty,N\right)  f_{1}\left(  k^{\prime}%
|\beta\rightarrow\infty,N\right)  , & k\neq k^{\prime},\\
1, & k=k^{\prime}<L,\\
0, & k=k^{\prime}>L,\\
\frac{N-N_{L}}{N_{L+1}-N_{L}}\frac{N-N_{L}-1}{N_{L+1}-N_{L}-1}, & k=k^{\prime
}=L.
\end{array}
\right.  \label{p2}%
\end{equation}

Using Eqs. (\ref{nf}) to (\ref{p2}), after performing integrations we arrive
at the following formulae for the Coulomb and electron-phonon energies
(\ref{t0}) to (\ref{t3}).

(i) The Coulomb energy for opposite spins is given by the expression%
\begin{align}
E_{C\uparrow\downarrow}\left(  N_{\uparrow},N_{\downarrow}\right)   &
=U\sqrt{\frac{2\omega}{\pi}}\sum_{k=0}^{L_{\uparrow}}\sum_{l=0}^{L_{\downarrow
}}\left(  -1\right)  ^{k+l}\binom{k-\frac{1}{2}}{k}\binom{k+l-\frac{1}{2}}%
{l}\nonumber\\
&  \times\left[  \binom{L_{\uparrow}+2}{k+3}+\frac{N_{\uparrow}-N_{L_{\uparrow
}}}{N_{L_{\uparrow}+1}-N_{L_{\uparrow}}}\binom{L_{\uparrow}+2}{k+2}\right]
\nonumber\\
&  \times\left[  \binom{L_{\downarrow}+2}{l+3}+\frac{N_{\downarrow
}-N_{L_{\downarrow}}}{N_{L_{\downarrow}+1}-N_{L_{\downarrow}}}\binom
{L_{\downarrow}+2}{l+2}\right]  . \label{ECa2}%
\end{align}

(ii) The Coulomb energy for parallel spins is%
\begin{align*}
E_{C\Vert}\left(  N\right)   &  =U\sqrt{\frac{\omega}{2\pi}}\sum_{k=0}^{L}%
\sum_{l=0}^{L}\left(  -1\right)  ^{k+l}\binom{k-\frac{1}{2}}{k}\binom
{k+l-\frac{1}{2}}{l}\\
&  \times\left[  \binom{L+2}{k+3}\binom{L+2}{l+3}+\frac{N-N_{L}}{N_{L+1}%
-N_{L}}\binom{L+2}{k+2}\binom{L+2}{l+3}\right. \\
&  +\frac{N-N_{L}}{N_{L+1}-N_{L}}\binom{L+2}{l+2}\binom{L+2}{k+3}\\
&  \left.  +\frac{N-N_{L}}{N_{L+1}-N_{L}}\frac{N-N_{L}-1}{N_{L+1}-N_{L}%
-1}\binom{L+2}{k+2}\binom{L+2}{l+2}\right] \\
&  -U\sqrt{\frac{\omega}{2\pi}}\sum_{k=0}^{L}\sum_{k^{\prime}=0}^{L}%
f_{2}\left(  k,k^{\prime}|\beta\rightarrow\infty,N\right) \\
&  \times\sum_{l=0}^{k_{<}}\sum_{j=0}^{l}\left(  -1\right)  ^{l+j}%
\binom{j-\frac{1}{2}}{j}\binom{k_{>}-k_{<}+l+j-\frac{1}{2}}{l+k_{>}-k_{<}%
}\binom{k_{>}+2}{k_{<}-l}\binom{2\left(  l+k_{>}-k_{<}\right)  }{l-j}.
\end{align*}

(iii) The electron-phonon energy for opposite spins is%
\begin{align}
E_{\alpha\uparrow\downarrow}\left(  N_{\uparrow},N_{\downarrow}\right)   &
=-2\alpha\sqrt{\frac{\omega}{\pi}}\int\limits_{0}^{\infty}d\tau e^{-\tau}%
\sum_{k=0}^{L_{\uparrow}}\sum_{l=0}^{L_{\downarrow}}\frac{\left(  -1\right)
^{k+l}\binom{k-\frac{1}{2}}{k}\binom{k+l-\frac{1}{2}}{l}}{\left[  2\omega
P\left(  \tau\right)  +1\right]  ^{k+l+\frac{1}{2}}}\nonumber\\
&  \times\left[  \binom{L_{\uparrow}+2}{k+3}+\frac{N_{\uparrow}-N_{L_{\uparrow
}}}{N_{L_{\uparrow}+1}-N_{L_{\uparrow}}}\binom{L_{\uparrow}+2}{k+2}\right]
\nonumber\\
&  \times\left[  \binom{L_{\downarrow}+2}{l+3}+\frac{N_{\downarrow
}-N_{L_{\downarrow}}}{N_{L_{\downarrow}+1}-N_{L_{\downarrow}}}\binom
{L_{\downarrow}+2}{l+2}\right]  \label{Epoppf}%
\end{align}
with the function%
\begin{equation}
P\left(  \tau\right)  \equiv\frac{1}{2\left(  N_{\uparrow}+N_{\downarrow
}\right)  }\left(  \frac{w^{2}}{v^{2}}\tau+\frac{v^{2}-w^{2}}{v^{3}}\left(
1-e^{-v\tau}\right)  -\frac{1-e^{-\omega\tau}}{\omega}\right)  . \label{P}%
\end{equation}

(iv) Finally, the electron-phonon energy for parallel spins takes the form%
\begin{equation}
E_{\alpha\parallel}\left(  N\right)  =-\frac{\sqrt{2}\alpha}{\pi}%
\int\limits_{0}^{\infty}d\tau e^{-\tau}E_{\alpha\parallel}\left(
N,\tau\right)  , \label{Ealpha}%
\end{equation}
where the time-dependent function $E_{\alpha\parallel}\left(  N,\tau\right)  $
is a sum of three terms:%
\begin{equation}
E_{\alpha\parallel}\left(  N,\tau\right)  =E_{\alpha\parallel}^{\left(
0\right)  }\left(  N,\tau\right)  +\frac{N-N_{L}}{N_{L+1}-N_{L}}%
E_{\alpha\parallel}^{\left(  1\right)  }\left(  N,\tau\right)  +\frac{N-N_{L}%
}{N_{L+1}-N_{L}}\frac{N-N_{L}-1}{N_{L+1}-N_{L}-1}E_{\alpha\parallel}^{\left(
2\right)  }\left(  N,\tau\right)  . \label{col}%
\end{equation}
The terms $E_{\alpha\parallel}^{\left(  j\right)  }\left(  N,\tau\right)  $
can be written down in two equivalent alternative forms. The first form is
relevant for the numerical calculation in the region of small and intermediate
values of $\left(  \omega\tau\right)  $,%
\begin{align}
E_{\alpha\parallel}^{\left(  0\right)  }\left(  N,\tau\right)   &
=\sqrt{\frac{\pi\omega}{2}}\left(  \sum_{j=0}^{L-1}\frac{\left[  2\left(
\cosh\omega\tau-1\right)  \right]  ^{j}\binom{L+2}{j+3}\binom{j-\frac{1}{2}%
}{j}}{\left[  2\omega P\left(  \tau\right)  +1-e^{-\omega\tau}\right]
^{j+\frac{1}{2}}}\right. \nonumber\\
&  -\sum_{j=1}^{L-1}\left(  e^{j\omega\tau}+e^{-j\omega\tau}\right)
\sum_{n=0}^{L-j-1}\sum_{m=0}^{n}\frac{\left(  -1\right)  ^{n+m}\binom
{L+2}{j+n+3}\binom{2\left(  j+n\right)  }{n-m}\binom{m-\frac{1}{2}}{m}%
\binom{j+n+m-\frac{1}{2}}{j+n}}{\left[  2\omega P\left(  \tau\right)
+1\right]  ^{j+n+m+\frac{1}{2}}}\nonumber\\
&  -\sum_{j=0}^{L-1}\sum_{n=0}^{j}\left(  -1\right)  ^{j+n}\frac{\binom
{2j}{j-n}\binom{L+2}{j+3}\binom{j-\frac{1}{2}}{j}\binom{j+n-\frac{1}{2}}{n}%
}{\left[  2\omega P\left(  \tau\right)  +1\right]  ^{j+n+\frac{1}{2}}%
}\nonumber\\
&  \left.  +\sum_{j=0}^{L-1}\sum_{n=0}^{L-1}\frac{\left(  -1\right)
^{j+n}\binom{L+2}{j+3}\binom{L+2}{n+3}\binom{j-\frac{1}{2}}{j}\binom
{j+n-\frac{1}{2}}{n}}{\left[  2\omega P\left(  \tau\right)  +1\right]
^{j+n+\frac{1}{2}}}\right)  , \label{o1}%
\end{align}%
\begin{align}
E_{\alpha\parallel}^{\left(  1\right)  }\left(  N,\tau\right)   &
=\sqrt{\frac{\pi\omega}{2}}\left(  \sum_{j=0}^{L}\frac{\left[  2\left(
\cosh\omega\tau-1\right)  \right]  ^{j}\binom{L+2}{j+2}\binom{j-\frac{1}{2}%
}{j}}{\left[  2\omega P\left(  \tau\right)  +1-e^{-\omega\tau}\right]
^{j+\frac{1}{2}}}\right. \nonumber\\
&  -\sum_{j=1}^{L}\left(  e^{j\omega\tau}+e^{-j\omega\tau}\right)  \sum
_{n=0}^{L-j}\sum_{m=0}^{n}\frac{\left(  -1\right)  ^{n+m}\binom{L+2}%
{j+n+2}\binom{2\left(  j+n\right)  }{n-m}\binom{m-\frac{1}{2}}{m}%
\binom{j+n+m-\frac{1}{2}}{j+n}}{\left[  2\omega P\left(  \tau\right)
+1\right]  ^{j+n+m+\frac{1}{2}}}\nonumber\\
&  \left.  +2\sum_{j=0}^{L-1}\sum_{n=0}^{L}\frac{\left(  -1\right)
^{j+n}\binom{L+2}{j+3}\binom{L+2}{n+2}\binom{j-\frac{1}{2}}{j}\binom
{j+n-\frac{1}{2}}{n}}{\left[  2\omega P\left(  \tau\right)  +1\right]
^{j+n+\frac{1}{2}}}\right)  , \label{o2}%
\end{align}%
\begin{align}
E_{\alpha\parallel}^{\left(  2\right)  }\left(  N,\tau\right)   &
=\sqrt{\frac{\pi\omega}{2}}\left(  -\sum_{j=0}^{L}\sum_{n=0}^{j}\left(
-1\right)  ^{j+n}\frac{\binom{2j}{j-n}\binom{L+2}{j+2}\binom{j-\frac{1}{2}}%
{j}\binom{j+n-\frac{1}{2}}{n}}{\left[  2\omega P\left(  \tau\right)
+1\right]  ^{j+n+\frac{1}{2}}}\right. \nonumber\\
&  \left.  +\sum_{j=0}^{L}\sum_{n=0}^{L}\frac{\left(  -1\right)  ^{j+n}%
\binom{L+2}{j+2}\binom{L+2}{n+2}\binom{j-\frac{1}{2}}{j}\binom{j+n-\frac{1}%
{2}}{n}}{\left[  2\omega P\left(  \tau\right)  +1\right]  ^{j+n+\frac{1}{2}}%
}\right)  . \label{o3}%
\end{align}
The second form is relevant for intermediate and large values of $\left(
\omega\tau\right)  $,%
\begin{align}
E_{\alpha\parallel}^{\left(  0\right)  }\left(  N,\tau\right)   &
=\sqrt{\frac{\pi\omega}{2}}\left(  \sum_{j=1}^{\infty}e^{-j\omega\tau}%
\sum_{n=0}^{L-1}\sum_{m=0}^{n}\frac{\left(  -1\right)  ^{n+m}\binom
{L+j+2}{n+j+3}\binom{2\left(  n+j\right)  }{n-m}\binom{m-\frac{1}{2}}{m}%
\binom{j+n+m-\frac{1}{2}}{j+n}}{\left[  2\omega P\left(  \tau\right)
+1\right]  ^{j+n+m+\frac{1}{2}}}\right. \nonumber\\
&  -\sum_{j=1}^{L-1}e^{-j\omega\tau}\sum_{n=0}^{L-j-1}\sum_{m=0}^{n}%
\frac{\left(  -1\right)  ^{n+m}\binom{L+2}{n+j+3}\binom{2\left(  n+j\right)
}{n-m}\binom{m-\frac{1}{2}}{m}\binom{j+n+m-\frac{1}{2}}{j+n}}{\left[  2\omega
P\left(  \tau\right)  +1\right]  ^{j+n+m+\frac{1}{2}}}\nonumber\\
&  \left.  +\sum_{n=0}^{L-1}\sum_{m=0}^{L-1}\frac{\left(  -1\right)
^{n+m}\binom{L+2}{n+3}\binom{L+2}{m+3}\binom{n-\frac{1}{2}}{n}\binom
{n+m-\frac{1}{2}}{m}}{\left[  2\omega P\left(  \tau\right)  +1\right]
^{n+m+\frac{1}{2}}}\right)  , \label{o4}%
\end{align}%
\begin{align}
E_{\alpha\parallel}^{\left(  1\right)  }\left(  N,\tau\right)   &
=\sqrt{\frac{\pi\omega}{2}}\left(  \sum_{j=0}^{\infty}e^{-j\omega\tau}%
\sum_{n=0}^{L}\sum_{m=0}^{n}\frac{\left(  -1\right)  ^{n+m}\binom
{L+j+2}{n+j+2}\binom{2\left(  n+j\right)  }{n-m}\binom{m-\frac{1}{2}}{m}%
\binom{j+n+m-\frac{1}{2}}{j+n}}{\left[  2\omega P\left(  \tau\right)
+1\right]  ^{j+n+m+\frac{1}{2}}}\right. \nonumber\\
&  -\sum_{j=1}^{L}e^{-j\omega\tau}\sum_{n=0}^{L-j}\sum_{m=0}^{n}\frac{\left(
-1\right)  ^{n+m}\binom{L+2}{n+j+2}\binom{2\left(  n+j\right)  }{n-m}%
\binom{m-\frac{1}{2}}{m}\binom{j+n+m-\frac{1}{2}}{j+n}}{\left[  2\omega
P\left(  \tau\right)  +1\right]  ^{j+n+m+\frac{1}{2}}}\nonumber\\
&  \left.  +2\sum_{n=0}^{L}\sum_{m=0}^{L-1}\frac{\left(  -1\right)
^{n+m}\binom{L+2}{n+2}\binom{L+2}{m+3}\binom{n-\frac{1}{2}}{n}\binom
{n+m-\frac{1}{2}}{m}}{\left[  2\omega P\left(  \tau\right)  +1\right]
^{n+m+\frac{1}{2}}}\right)  . \label{o5}%
\end{align}

\newpage%

\begin{figure}
[ptbh]
\begin{center}
\includegraphics[
height=4.7011in,
width=5.4812in
]%
{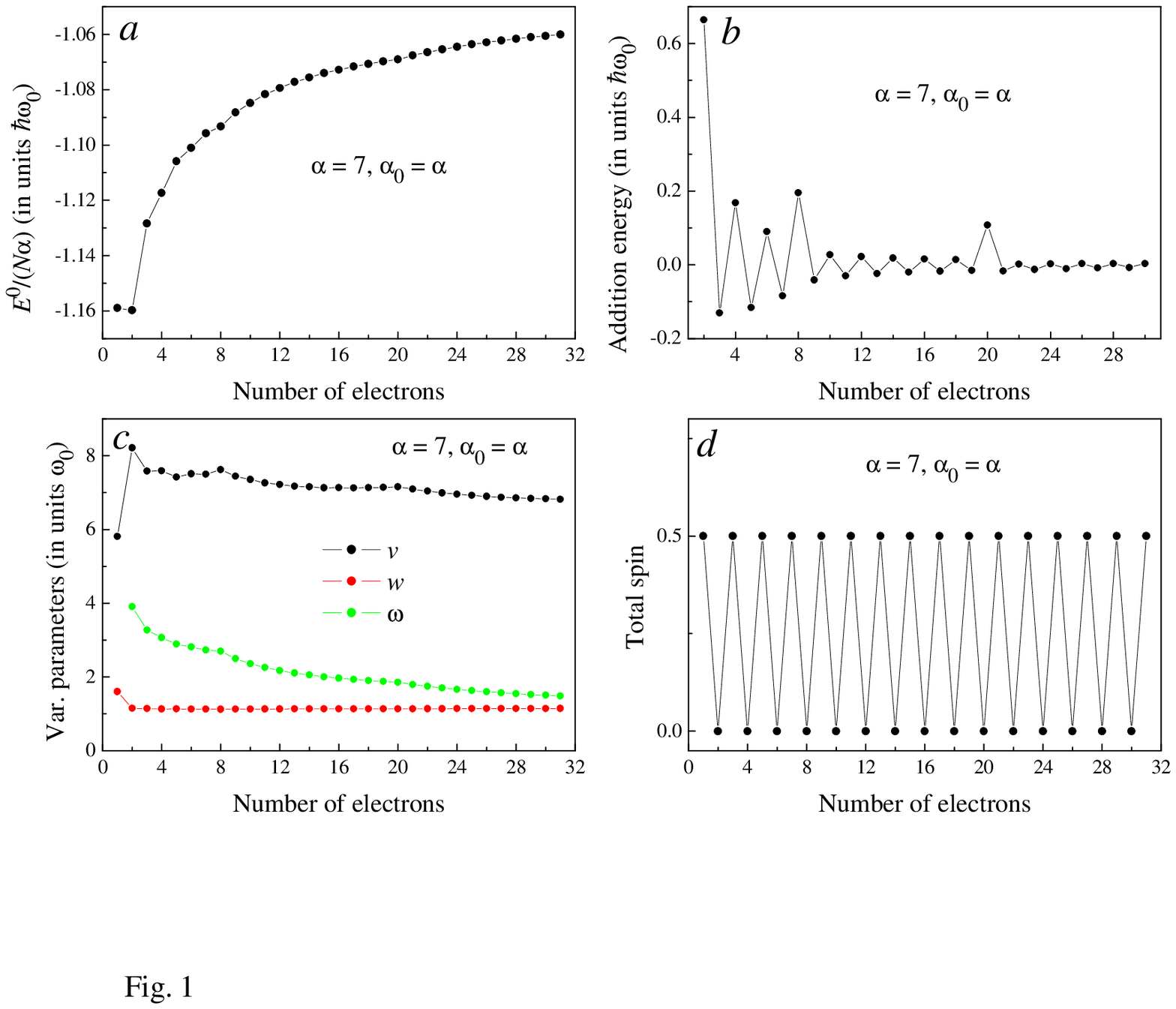}%
\end{center}
\end{figure}

Fig. 1. (Color online) The ground-state energy per polaron (\emph{a}), the
addition energy (\emph{b}), the optimal values of variational parameters
(\emph{c}) and the total spin (\emph{d}) as a function of $N$ for a
translation invariant $N$-polaron system with $\alpha=7$, $\alpha_{0}=\alpha$.

\newpage%

\begin{figure}
[ptbh]
\begin{center}
\includegraphics[
height=5.047in,
width=5.4051in
]%
{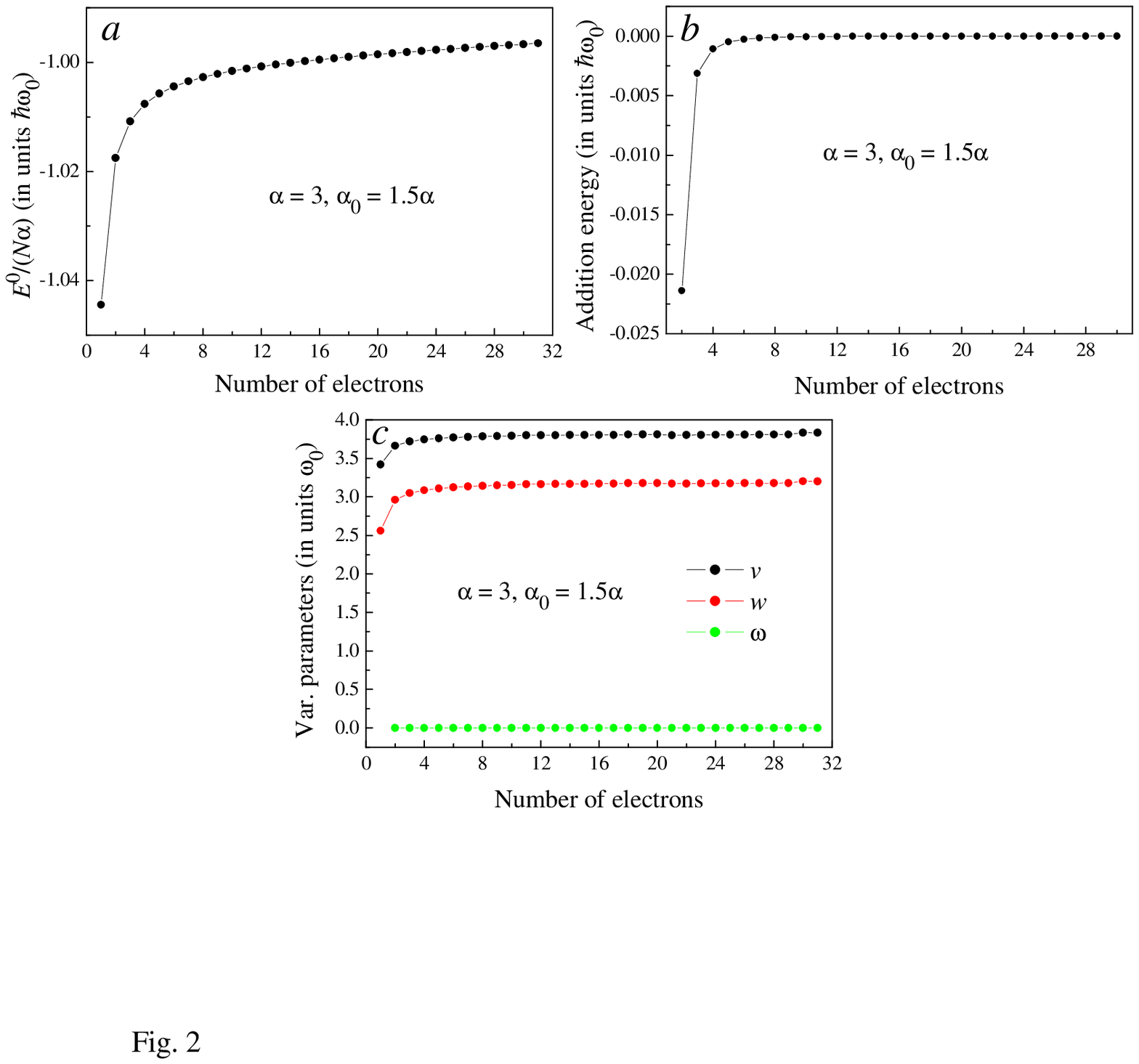}%
\end{center}
\end{figure}

Fig. 2. (Color online) The ground-state energy per polaron (\emph{a}), the
addition energy (\emph{b}), and the optimal values of variational parameters
(\emph{c}) as a function of $N$ for a translation invariant $N$-polaron system
with $\alpha=3$, $\alpha_{0}=1.5\alpha$.

\newpage%

\begin{figure}
[ptbh]
\begin{center}
\includegraphics[
height=5.047in,
width=5.4405in
]%
{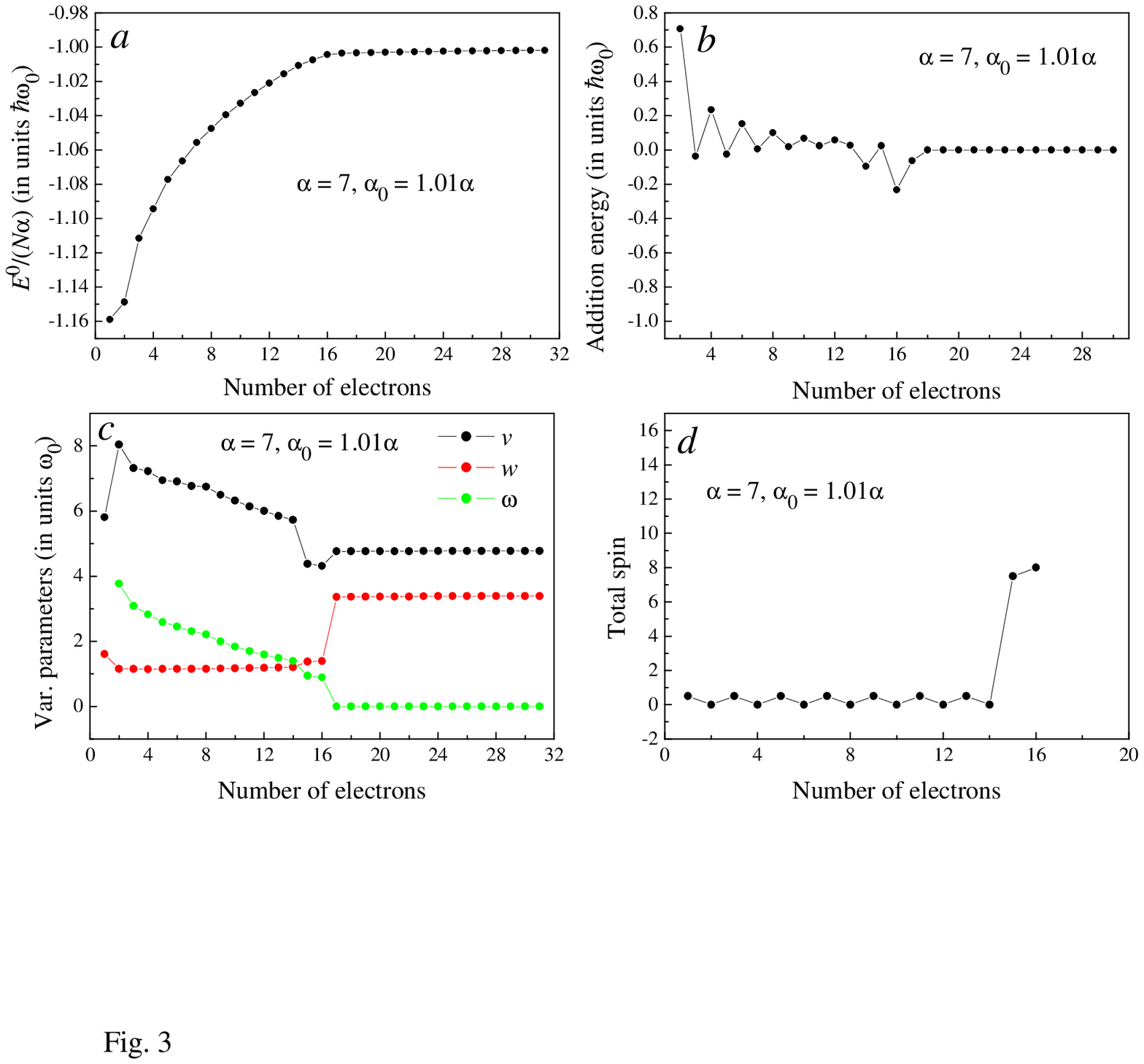}%
\end{center}
\end{figure}

Fig. 3. (Color online) The ground-state energy per polaron (\emph{a}), the
addition energy (\emph{b}), the optimal values of variational parameters
(\emph{c}) and the total spin (\emph{d}) as a function of $N$ for a
translation invariant $N$-polaron system with $\alpha=7$, $\alpha
_{0}=1.01\alpha$.

\newpage%

\begin{figure}
[ptbh]
\begin{center}
\includegraphics[
height=5.3774in,
width=4.4771in
]%
{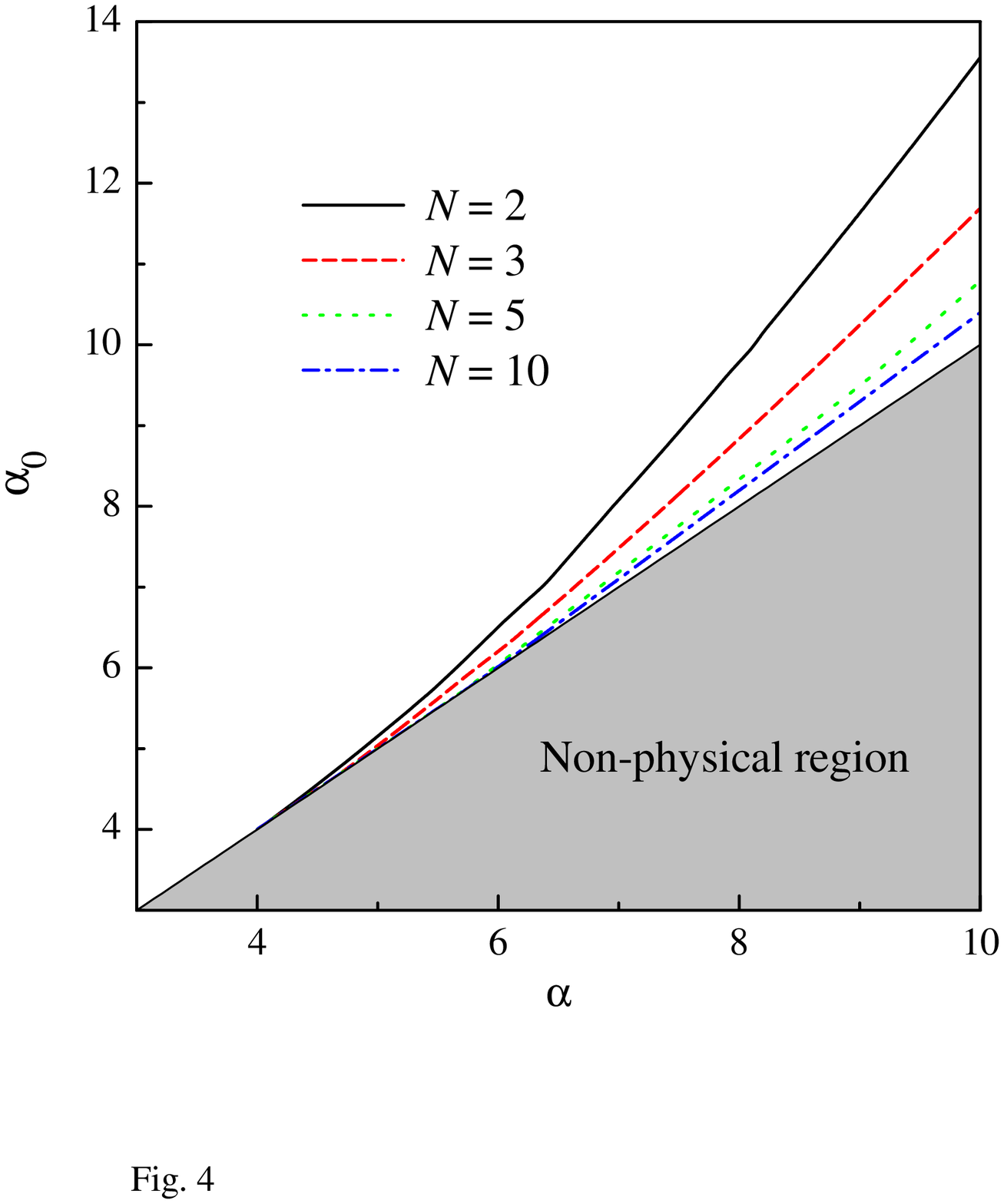}%
\end{center}
\end{figure}

Fig. 4. (Color online) The \textquotedblleft phase diagrams\textquotedblright%
\ of a translation invariant $N$-polaron system. The grey area is the
non-physical region, for which $\alpha>\alpha_{0}$. The stability region for
each number of electrons is determined by the equation $\alpha_{c}%
<\alpha<\alpha_{0}$.

\newpage%

\begin{figure}
[ptbh]
\begin{center}
\includegraphics[
height=6.7715in,
width=3.7005in
]%
{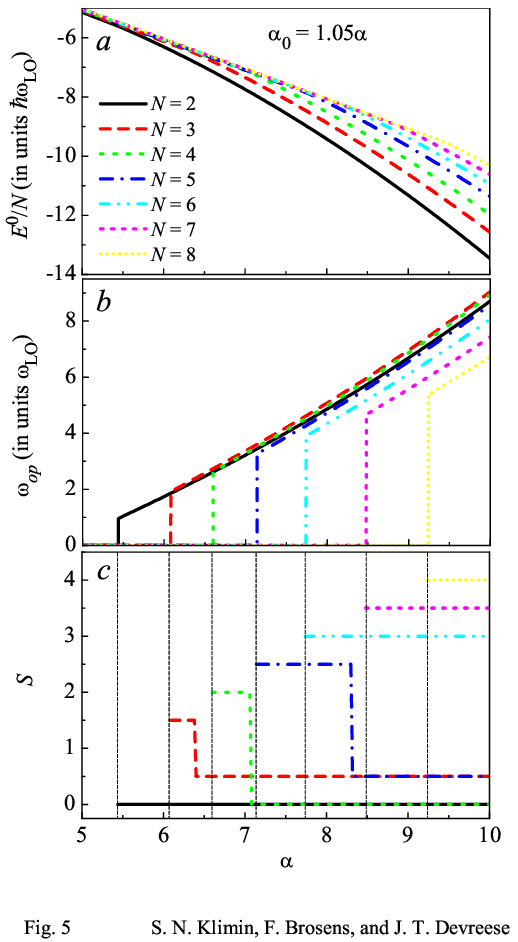}%
\end{center}
\end{figure}

Fig. 5. (Color online) The ground-state energy per particle (\emph{a}), the
optimal value $\omega_{op}$ of the confinement frequency (\emph{b}), and the
total spin (\emph{c}) of a translation invariant $N$-polaron system as a
function of the coupling strength $\alpha$ for $\alpha_{0}/\alpha=0.5$. The
vertical dashed lines in the panel \emph{c} indicate the critical values
$\alpha_{c}$ separating the regimes of $\alpha>\alpha_{c}$, where the
multipolaron ground state with $\omega_{op}\neq0$ exists, and $\alpha
<\alpha_{c}$, where $\omega_{op}=0$.

\end{document}